\documentclass[twocolumn]{aastex631}

\usepackage{newtxtext,newtxmath,color}
\usepackage{amsmath, graphics, bm, graphicx}
\allowdisplaybreaks

\newcommand{\mathsym}[1]{{}}
\newcommand{\unicode}[1]{{}}

\newcommand{\kg}{\kappa}

\newcommand{\Om}{\Omega}
\newcommand{\om}{\omega}

\newcommand{\pd}{\partial}

\newcommand{\der}{{\rm d}}

\newcommand{\e}{{\rm e}}

\newcommand{\Ms}{M_\star}
\newcommand{\Rs}{R_\star}

\newcommand{\sg}{\sigma}

\newcommand{\ro}{{\rm orb}}

\newcommand{\Mjup}{M_{\rm Jup}}

\newcommand{\Msun}{M_\odot}

\newcommand{\Oms}{\Omega_\star}

\newcommand{\Mp}{M_{\rm p}}
\newcommand{\bJs}{{\bm J}_\star}
\newcommand{\hjs}{{\bm {\hat j}}_\star}
\newcommand{\bJo}{{\bm J}_{\rm orb}}
\newcommand{\hjo}{{\bm {\hat j}}_{\rm orb}}

\newcommand{\bcdot}{{\bm \cdot}}

\newcommand{\s}{{\rm sec}}

\newcommand{\be}{\begin{equation}}
\newcommand{\ee}{\end{equation}}

\begin{document}

\title{Spin and Obliquity Evolution of Hot Jupiter Hosts from Resonance Locks}

\author[0000-0002-9849-5886]{J. J. Zanazzi}\thanks{51 Pegasi b fellow, email: jzanazzi@berkeley.edu}
\affiliation{
Astronomy Department, Theoretical Astrophysics Center, and Center for Integrative Planetary Science, University of California, Berkeley, \\
Berkeley, CA 94720, USA \\
}

\author[0000-0002-6246-2310]{Eugene Chiang}
\affiliation{
Astronomy Department, Theoretical Astrophysics Center, and Center for Integrative Planetary Science, University of California, Berkeley, \\
Berkeley, CA 94720, USA \\
}
\affiliation{
Department of Earth and Planetary Science, University of California, Berkeley, CA 94720, USA
}



\begin{abstract}
When a hot Jupiter orbits a star whose effective temperature exceeds $\sim$6100 K, its orbit normal tends to be misaligned with the stellar spin axis. Cooler stars typically have smaller obliquities, which may have been damped by hot Jupiters in resonance lock with axisymmetric stellar gravity modes (azimuthal number $m=0$). Here we allow for resonance locks with non-axisymmetric modes, which affect both stellar obliquity and spin frequency. Obliquities damp for all modes ($-2 \leq m \leq 2$). Stars spin up for $m > 0$, and spin down for $m < 0$. We carry out a population synthesis that assumes hot Jupiters form misaligned around both cool and hot stars, and subsequently lock onto modes whose $m$-values yield the highest mode energies for given starting obliquities. Core hydrogen burning enables hot Jupiters to torque low-mass stars, but not high-mass stars, into spin-orbit alignment. Resonance locking plus stellar spin-down from magnetic braking largely reproduces observed obliquities and stellar rotation rates and how they trend with stellar effective temperature and orbital separation. The possible suppression of resonance locking by non-linear dissipation of gravity waves remains an outstanding issue.
\end{abstract}


\section{Introduction}

The angle between the orbit normal of a hot Jupiter, and the stellar spin axis of its host star, can vary depending on the host's effective temperature: cool stars tend to be aligned, while hot stars are often misaligned \citep[e.g.][]{Winn+(2010), Winn+(2017), Schlaufman(2010), Albrecht+(2012), Albrecht+(2021), MunozPerets(2018), HamerSchlaufman(2022), Rice+(2022), Rice+(2022b), Siegel+(2023), Knudstrup+(2024), Wang+(2024)}.  Previous attempts to explain this trend posit tidal damping is more efficient in the convective envelopes of stars below the `Kraft break' (effective temperatures $\le 6100 \ {\rm K}$).  The convective envelope mechanisms, however, have shortcomings: equilibrium tides have difficulty aligning orbits before the hot Jupiter is engulfed \citep[e.g.][]{BarkerOgilvie(2009), Winn+(2010), Dawson(2014)}, and inertial waves have trouble damping retrograde obliquities \citep[e.g.][]{Lai(2012), RogersLin(2013), ValsecchiRasio(2014), Xue+(2014), LiWinn(2016), LinOgilvie(2017), DamianiMathis(2018), Anderson+(2021), SpaldingWinn(2022)}.  

More recently, tidal `resonance locking' has been suggested to explain the obliquity and spin evolution of hot Jupiter systems.  Tidal forcing and dissipation increase when a natural oscillation of the host star is resonantly excited by the planet, i.e.~when the star's oscillation frequency lies close to a harmonic of the planet's orbital frequency.  Such resonances can be maintained even as the star's internal structure and oscillation frequencies change \citep[e.g.][]{WitteSavonije(1999), WitteSavonije(2001), Savonije(2008)}.  For Sun-like stars, low-frequency gravity modes (g-modes), which propagate only in radiative (stably stratified) regions, might be excited by hot Jupiters.  \cite{MaFuller(2021)} suggested that the faster rotation rates of hot Jupiter hosts having cooler effective temperatures \citep[e.g.][]{Penev+(2018)} might be explained by tidal spin-up from resonance locking, though they urged caution in this interpretation because of the potential saturation of resonances by a parametric instability (see Appendix B of \citealt{Zanazzi+(2024)} for a discussion, and also the last paragraph of the present paper).  
\cite{Zanazzi+(2024)} established that hydrogen burning causes g-modes in the stably stratified cores of cool stars to increase their frequencies by order-unity factors on the main sequence, while g-modes confined to the radiative envelopes of hot stars remain largely unaffected. \cite{Zanazzi+(2024)} further showed that resonance locking to zonal, axisymmetric g-modes (symmetric about the stellar spin axis) preferentially damps the obliquities of low-mass hot Jupiter hosts, while leaving the obliquities of high-mass hosts comparatively unaltered. \cite{Millholland+(subm)} followed this work by showing how the orbital period distribution of hot Jupiters around cool stars is consistent with migration by resonance locking, unlike the orbits of hot Jupiters around hot stars.

The goal of the present work is to extend \cite{Zanazzi+(2024)}, whose analysis was largely restricted to axisymmetric g-modes, and explore how spin and obliquity evolve in tandem from resonance locks with non-axisymmetric g-modes.   Section~\ref{sec:StarModel} presents stellar evolution models used to calculate the properties of g-modes, as well as a prescription of stellar magnetic braking, which spins stars down (more so for cool stars than hot stars).  Section~\ref{sec:ResLock} considers how hot Jupiters may lock onto non-axisymmetric stellar modes, and the consequences for stellar obliquity and spin. Section~\ref{sec:PopSynth} carries out a population synthesis that makes direct comparison between simulated and observed obliquities and rotation rates of cool and hot stars. Section~\ref{sec:SummDisc} summarizes, and discusses future directions.

\section{Evolution of hot Jupiter host stars}
\label{sec:StarModel}

Nuclear burning changes the internal structure of a star and by extension its gravity-mode (g-mode) oscillation frequencies. At the same time, magnetic braking slows a star's rotation.  We describe here  stellar models incorporating both effects, for later use in resonance locking calculations.  We consider a star with mass $\Ms$, radius $\Rs$, spin frequency $\Oms$, period $P_\star = 2\pi/\Oms$, moment of inertia $I_\star = \kappa_\star M_\star R_\star^2$, and spin angular momentum ${\bm J}_\star = I_\star \Omega_\star \hjs$, with $\hjs$ the unit vector parallel to the stellar spin axis.  The dimensionless moment of inertia constant $\kappa_\star$ equals
\be
\kg_\star = \frac{8\pi}{3\Ms \Rs^2} \int_0^{\Rs} \rho r^4 \der r.
\label{eq:kgs}
\ee

\subsection{Gravity-mode frequency evolution}

\begin{figure}
\centering
\includegraphics[width=\linewidth]{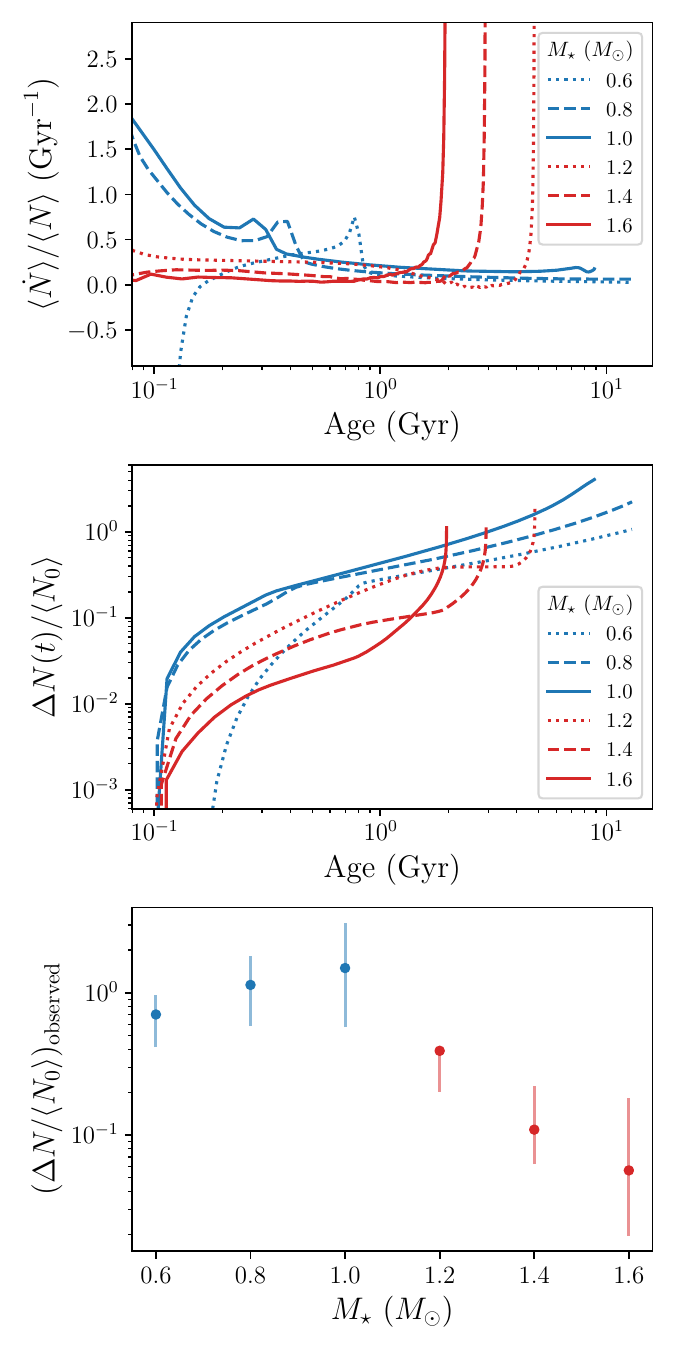}
\caption{
Rate of Brunt-V\"ais\"al\"a frequency evolution (top panel, eq.~\ref{eq:N_ave}) and accumulated frequency change (middle panel, eq.~\ref{eq:Porb_brunt}) vs.~main-sequence age, for stars of different masses. Although high-mass stars (red curves) increase their frequencies dramatically toward the end of their hydrogen-burning lives (as their cores switch from being unstably to stably stratified), the speed-up phase is short-lived and unlikely to be observed. Sampled over the entirety of their main-sequence lives, $\Delta N/ \langle N_0 \rangle$ is more likely to be higher for low-mass stars, whose cores remain stably stratified throughout 
(bottom panel showing median and $\pm 1 \sigma$ intervals of $(\Delta N/\langle N_0 \rangle)_{\rm observed}$, calculated by evaluating $\Delta N/\langle N_0 \rangle$ at 1000 uniformly spaced times over the interval $[t_0, t_{\rm max}]$, with $t_{\rm max} = \min(t_{\rm MS}, 12 \, {\rm Gyr})$ and $t_{\rm MS}$ is the age where core hydrogen burning ceases. The upper $+1\sigma$ interval for $M_\star = 1.2 M_\odot$ is too small to see).  Compare with Fig.~6 of \cite{Zanazzi+(2024)}; the calculations here start at an earlier time of $\sim$0.1 Gyr, when hot Jupiters may have formed from high-eccentricity migration.
 \label{fig:mode_ev}
}
\end{figure}

In the low-frequency limit, g-mode frequencies in the host star's rotating frame are given by
\be
\om \simeq \frac{\sqrt{\ell(\ell+1)}}{\pi n} \langle N \rangle,
\label{eq:om_ag}
\ee
where $n \gg 1$ is the number of radial nodes, and the radial average of the Brunt-V\"ais\"al\"a frequency is given by
\be
\langle N \rangle = \int_{\rm rad} \frac{N}{r} \der r,
\label{eq:N_ave}
\ee
with the integral taken over the stably stratified radiative zone ($N^2>0$).  Hydrogen burning increases the mean molecular weight and by extension the stratification.  This increases $\langle N \rangle$, more so for cool stars with stably stratified cores.

Following \cite{Zanazzi+(2024)}, we calculate the evolution of $\langle N \rangle$ using \texttt{MESA} stellar structure models up to an age 
\begin{equation}
\frac{\Delta N(t)}{\langle N_0 \rangle} \equiv \frac{1}{\langle N_0 \rangle} \int_{t_0}^t \max\left( \langle \dot N \rangle, 0 \right) \der t,
    \label{eq:Porb_brunt}
\end{equation}
which omits changes in $\Delta N$ during intervals when $\langle \dot N \rangle < 0$ and resonance locks momentarily cease (e.g. \citealt{Fuller(2017), ZanazziWu(2021), MaFuller(2021)}). Figure~\ref{fig:mode_ev} shows how the Brunt-V\"ais\"al\"a frequency varies with time and stellar mass (compare with Fig.~6 of \citealt{Zanazzi+(2024)}). Generally $\langle \dot N \rangle > 0$ as nuclear burning causes stars to become more stratified. An exception occurs for our 0.6 $M_\odot$ model at $t \lesssim 0.2$ Gyr, when  ${}^3{\rm He}$ and ${}^{12}{\rm C}$ burning renders the core convectively unstable \citep[e.g.][]{Iben(1965), Gough(1980)} and $\langle \dot N \rangle < 0$. For this model, once the core ${}^3$He and ${}^{12}$C are depleted, the $p$-$p$ chain dominates energy generation, the core becomes radiative, and $\langle \dot N \rangle$ spikes. Similar spikes can be seen in the 0.8 and $1.0 M_\odot$ models.

\subsection{Spin-down from magnetic braking}\label{subsec:spin_down}

\begin{figure}
\centering
\includegraphics[width=\linewidth]{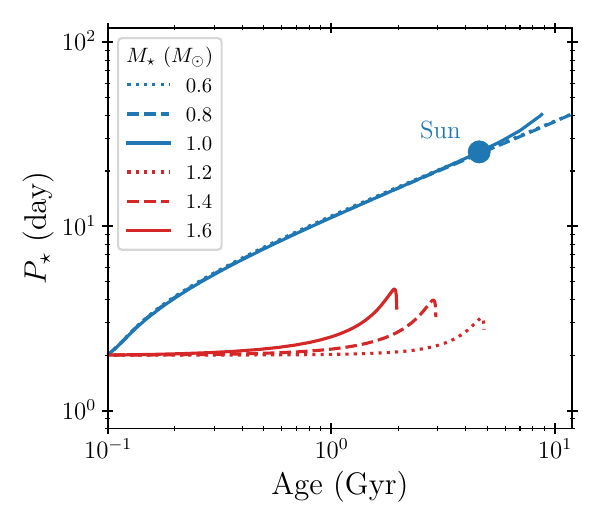}
\caption{
Evolution of a star's spin period $P_\star$, accounting for its changing moment of inertia (eq.~\ref{eq:kgs}) and magnetic braking (eq.~\ref{eq:dJsdt_wind}).  Low-mass stars ($\Ms < 1.2 \ \Msun$) spin down from magnetic braking, while high-mass stars ($\Ms \ge 1.2 \ \Msun$) spin down from radius expansion on the main sequence.  Our magnetic braking prescription is such that a $1.0 M_\odot$ model reproduces the Sun's current rotation period of $25.4$ days at an age of 4.57 Gyr. For this figure only, we initialize all stellar models with $P_\star = 2 \ {\rm days}$ at an age of 
100 Myr.
\label{fig:Spinwind}
}
\end{figure}

Stars lose angular momentum through a magnetized wind. Following, e.g., \citet{Skumanich(1972)} and \citet{Krishnamurthi+(1997)}, we prescribe this angular momentum loss to follow
\begin{equation}
    \left. \frac{\der \bJs}{\der t} \right|_{\rm wind} = -K_{\rm wind}  \min(\Om_{\rm sat}, \Om_\star)^2 \bJs
    \label{eq:dJsdt_wind}
\end{equation}
where $\Om_{\rm sat}$ is a saturation rotation rate, and the constant
\begin{equation}
    K_{\rm wind} = \bar K K_\odot \left( \frac{\Ms}{M_\odot} \right)^{-1/2} \left( \frac{\Rs}{R_\odot} \right)^{1/2}
\end{equation}
with $K_\odot$ tuned to reproduce the Sun's rotation rate of $\Omega_\odot = (2\pi)/(25.4 \ \der)$ at an age of $t_\odot = 4.57 \ {\rm Gyr}$:
\begin{equation}
    K_\odot \simeq \frac{1}{2 \Om_\odot^2 t_\odot} = 4.9 \times 10^{-12} \ \der.
\end{equation}
We set $\bar K = 1$ when $\Ms < 1.2 \ \Msun$, and $\bar K = 0.2$ when $\Ms \ge 1.2 \ \Msun$, to model weaker magnetic braking for more massive stars with radiative envelopes \citep[e.g.][]{Kraft(1967), Amard+(2019), Gossage+(2021)}. We fix $\Omega_{\rm sat} = 10 \Omega_\odot$. Our model reproduces the Sun's current rotation period, and predicts K and G stars to spin down substantially as they age, by contrast to F and A stars (Fig.~\ref{fig:Spinwind}).

\section{Spin and obliquity evolution during resonance locks}
\label{sec:ResLock}

\begin{figure}
\centering
\includegraphics[width=\linewidth]{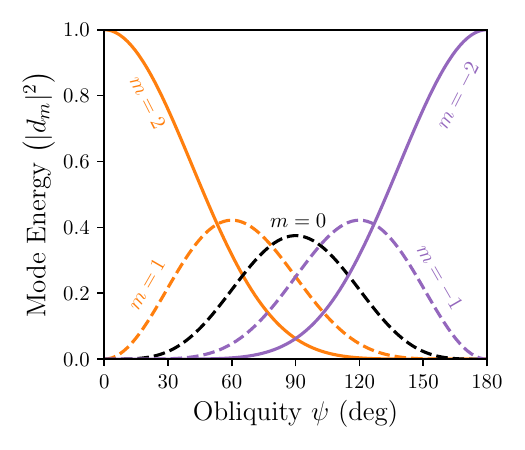}
\caption{
Mode energy amplitude as a function of obliquity (eqs.~\ref{eq:E_mode} and~\ref{eq:Wigner_d}).  When $\psi=0$, only $m=2$ has non-zero energy.  As $\psi$ increases, successively lower $m<2$ modes are excited.
\label{fig:d2_amp}
}
\end{figure}

\begin{figure*}
\centering
\includegraphics[width=\linewidth]{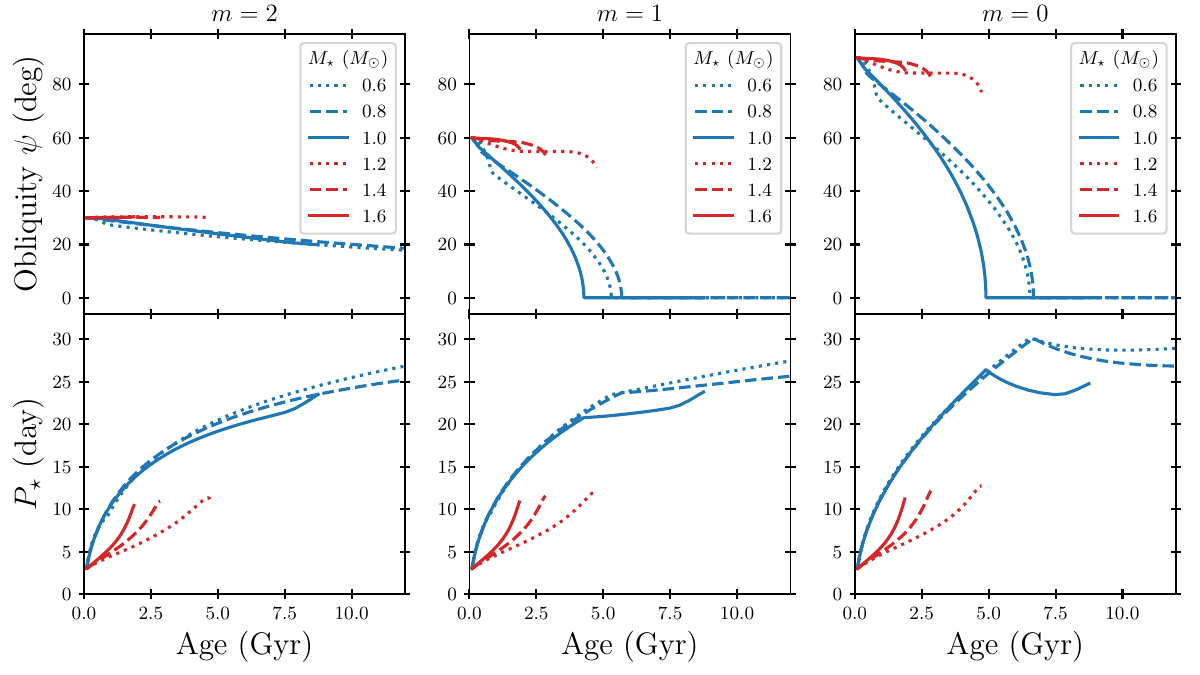}
\caption{
Orbital evolution in resonance lock  (eqs.~\ref{eq:dota_lock}-\ref{eq:dotpsi_lock}) for different initial obliquities ($\psi_0 = 30^\circ, 60^\circ, 90^\circ$) which lock to modes with different $m$ values as indicated (see discussion around eq.~\ref{eq:E_mode} and Fig.~\ref{fig:d2_amp}). For all curves, $\Mp = 1 \ \Mjup$, with initial conditions 
$(a/\Rs)_0 = 10$ and 
$P_{\star,0} = 3 \ {\rm days}$ at
$t_0 = 0.1$ Gyr.  
\label{fig:SpinObl_prog}
}
\end{figure*}

\begin{figure*}
\centering
\includegraphics[width=0.7\linewidth]{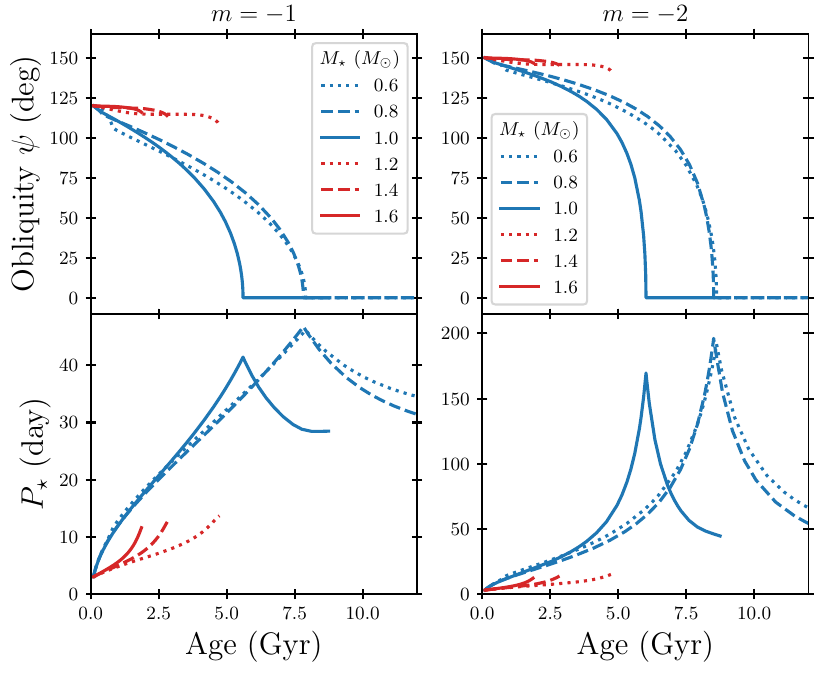}
\caption{
Same as Fig.~\ref{fig:SpinObl_prog}, except for initially retrograde obliquities ($\psi_0 = 120^\circ, 150^\circ$).  The $y$-axis scale differs between the left and right plots for $P_\star$.
\label{fig:SpinObl_retro}
}
\end{figure*}

Following \citet{Zanazzi+(2024)}, we present a general set of equations for the obliquity and orbit evolution of hot Jupiter systems that are resonantly locked.  The star is orbited by a planet of mass $\Mp \ll \Ms$, on a circular orbit ($e=0$) of semi-major axis $a$, mean-motion $\Om = \sqrt{G \Ms/a^3}$,  energy $E_\ro = -G \Ms \Mp/(2a)$, and angular momentum $\bJo = - (2 E_\ro/\Om)\hjo$, where $G$ is the gravitational constant and $\hjo$ is the unit orbit normal. The stellar obliquity $\psi$ is the angle between $\bJs$ and $\bJo$ ($\cos \psi = \hjo \bcdot \hjs$).

In resonance lock, a match between an oscillation frequency $\sigma$ in the star's inertial frame, and a harmonic of the orbital frequency $k \Omega$, is maintained by stellar evolution:
\begin{equation}
    \sigma = \omega + m \Omega_\star = k \Omega 
    \label{eq:res_lock}
\end{equation}
where $\omega$ is the mode frequency in the star's rotating frame, and $m$ is the mode's azimuthal number.  The stellar spin and planet orbital frequencies adjust to maintain resonance as $\omega$ increases from the changing internal structure of the star:
\begin{equation}
    t_{\rm ev}^{-1} = \frac{\dot \omega}{\omega} = \eta \frac{\dot E_{\rm orb}}{E_{\rm orb}} 
\end{equation}
where
\be
\eta = \frac{3}{2} - \frac{m^2 B}{2 k^2} \left( \frac{\Mp a^2}{\kg_\star M_\star \Rs^2} \right) 
\,.
\label{eq:eta}
\ee
For $g$-modes which satisfy $\Oms \ll \sg$,
\be
B = \frac{1}{m} \frac{\pd \sg}{\pd \Oms} \simeq 1 - \frac{1}{\ell(\ell+1)}
\ee
where $\ell$ is the angular degree of the mode. The parameter $\eta$ is typically positive for hot Jupiter systems; for example, for our solar-mass \texttt{MESA} model with $a/\Rs = 10$ and $m=2$, $\eta > 0$  
when $\Mp \lesssim 4 \ \Mjup$. Because tidal dissipation shrinks the orbit, resonance lock requires $t_{\rm ev}^{-1} > 0$ ($\dot{\omega} > 0$) for $\eta > 0$ \citep{FullerLai(2012), Fuller(2017), ZanazziWu(2021)}.  
From here on, we fix $\{k, \ell\} = \{2, 2\}$ as this term of the tidal forcing potential dominates when the companion has a circular orbit \citep{Zanazzi+(2024)}.

Dissipation of the oscillation induces a tidal lag, resulting in a mutual torque between star and companion.  As $\omega$ evolves, the torque modifies the orbit and the stellar spin to maintain resonance:
\begin{align}
    &\frac{1}{a} \frac{\der a}{\der t} = - \frac{1}{\eta t_{\rm ev}}
    \label{eq:dota_lock}\\
    &\frac{\der J_\star}{\der t} = \frac{m}{4} \frac{J_{\rm orb}}{\eta t_{\rm ev}} + \frac{\der J_\star}{\der t} \bigg|_{\rm wind}
    \label{eq:dotJs_lock}\\
    &\frac{\der \psi}{\der t} = - \frac{1}{4} \left[ \tau_m \left( \frac{J_{\rm orb}}{J_\star} + \cos \psi \right) - m \sin\psi \right] \frac{1}{\eta t_{\rm ev}},
    \label{eq:dotpsi_lock}
\end{align}
where $\tau_m = 2 \Omega T_x/\dot E$ is a dimensionless measure of the torque $T_x$ perpendicular to $\hjs$, normalized by the energy dissipation rate $\dot E = - \dot E_{\rm orb} > 0$.
For $|m| \le \ell = 2$, 
\begin{equation}    
    \tau_{\pm 2} = \frac{2 \sin \psi}{1 \pm \cos \psi},
    \hspace{2mm}
    \tau_{\pm 1} = \frac{2 \mp \cos \psi}{\sin \psi},
    \hspace{2mm}
    \tau_0 = \frac{2}{\sin \psi}.
    \label{eq:tau_vals}
\end{equation}
Notice $\der J_\star/\der t$ includes the magnetic braking torque $\der J_\star/\der t|_{\rm wind}$ (eq.~\ref{eq:dJsdt_wind}).  Because resonance locks cannot be maintained when either $t_{\rm ev}^{-1} \propto \langle \dot N \rangle <0$ 
or $\eta < 0$ (but not both),s
we set $t_{\rm ev}^{-1} = 0$ in equations \eqref{eq:dota_lock}--\eqref{eq:dotpsi_lock} under such circumstances.  We also switch to an $m=0$ lock if, for the initial $m \ne 0$ lock, $\eta < 0$, $\psi>0$, and $\langle \dot N \rangle > 0$; an $m=0$ mode has $\eta = 3/2 > 0$.

Near resonance, when $\omega$ is close to the tidal forcing frequency in the star's rotating frame $\omega_m = 2 \Omega - m \Oms$, the energies of modes with different $m$ depend on obliquity:
\begin{align}
    \frac{E_{\rm mode}}{E_\star} &= \frac{\omega^2}{(\omega - \omega_{m})^2 + \gamma^2} \left( \frac{\Mp}{\Ms} \right)^2 \nonumber \\
    &\times \left( \frac{\Rs}{a} \right)^6 \left( \frac{3\pi}{10} \right) |I|^2 |d_m(\psi)|^2,
    \label{eq:E_mode}
\end{align}
where $E_\star = G M_\star^2/R_\star$ is the star's binding energy, $\gamma$ is the mode damping rate, the mode overlap integral $|I|$ does not depend on $m$ or $\psi$, and the Wigner-$d$ coefficients are given by
\begin{align}
    d_{\pm 2} &= \frac{1}{4}(1 \pm \cos \psi)^2 
    \nonumber \\
    d_{\pm 1} &= \frac{1}{2} \sin \psi (1 \pm \cos \psi)
    \nonumber \\
    d_{0} &= \sqrt{ \frac{3}{8} } \sin^2 \psi.
    \label{eq:Wigner_d}
\end{align}
The dependence of $E_{\rm mode}$ on obliquity is set entirely by $|d_m|^2$ (Fig.~\ref{fig:d2_amp}).  For aligned orbits, only $m=2$ oscillations are excited, as expected \citep[e.g.][]{MaFuller(2021)}.  As $\psi$ increases from $0^\circ$, 
successively smaller (more negative) $m$ modes have higher energies.

In calculating the orbital evolution from resonance locks, we select, for a given initial obliquity $\psi_0$, the value of $m$ which maximizes the mode energy~\eqref{eq:E_mode}. 
The rationale here (and it is not more than a plausibility argument) is that the mode with the highest energy has the highest energy dissipation rate ($\dot E_{\rm mode} = 2 \gamma E_{\rm mode}$), and resonance locking depends on an energy dissipation rate that varies strongly with forcing frequency.  We then integrate equations~\eqref{eq:dota_lock}-\eqref{eq:dotpsi_lock} for that $m$ until $\psi=0^\circ$, at which point we switch to $m=2$.  Figures~\ref{fig:SpinObl_prog}-\ref{fig:SpinObl_retro} display a few sample integrations, taking $t_{\rm ev}^{-1}$ values from Figure~\ref{fig:mode_ev} and advancing the hot Jupiter's orbit starting from $t_0 = 0.1 \ {\rm Gyr}$.  With the exception of stars that start locked in $m=2$, all low-mass stars ($\Ms < 1.2 \ \Msun$) have their obliquities $\psi$ damped to $0^\circ$ in roughly half the main-sequence lifetime.  High-mass stars ($\Ms \ge 1.2 \ \Msun$) have obliquities which remain close to their initial values. 
Obliquities decay faster for lower-mass stars because they have higher $t_{\rm ev}^{-1}$ (from their hydrogen-burning radiative cores; Fig.~\ref{fig:mode_ev}), and because magnetic braking (from their convective envelopes) lowers $J_\star$, increasing $\dot \psi$  (eq.~\ref{eq:dotpsi_lock}).  
These findings largely reinforce those of our earlier work \citep{Zanazzi+(2024)}.

Magnetic braking increases stellar spin periods $P_\star$, more so for low-mass stars and their magnetized convective envelopes. High-mass stars also slow down their spins, mostly from angular momentum conservation as their radii increase on the main sequence. Tidal torques from resonance locking can modify the spin evolution for low-mass stars. For $m<0$, tides enhance spin-down; $P_\star$ can exceed 100 days in an $m=-2$ lock. The spin evolution abruptly changes once $\psi$ reaches $0^\circ$ and the lock is switched to $m=2$. Since  $\dot J_\star \propto m/t_{\rm ev}$, switching to $m=2$ introduces a spin-up torque that competes with and can defeat magnetic braking.

\section{Population Synthesis}
\label{sec:PopSynth}

We have shown that hot Jupiter resonance locks damp obliquities irrespective of $m$, increase the star's rotation rate when $m>0$, and decrease the rotation rate when $m<0$.  Here we carry out a population synthesis to more directly compare our theory to observations. We first examine in section \ref{subsec:rot_vel} what our magnetic braking model predicts for the rotation periods and velocities of field main-sequence stars. In section \ref{subsec:oblique}, we explore the predictions of our full hot Jupiter tidal evolution model.

\subsection{Rotational velocities from magnetic braking}\label{subsec:rot_vel}

\begin{figure}[ht!]
\centering
\includegraphics[width=\linewidth]{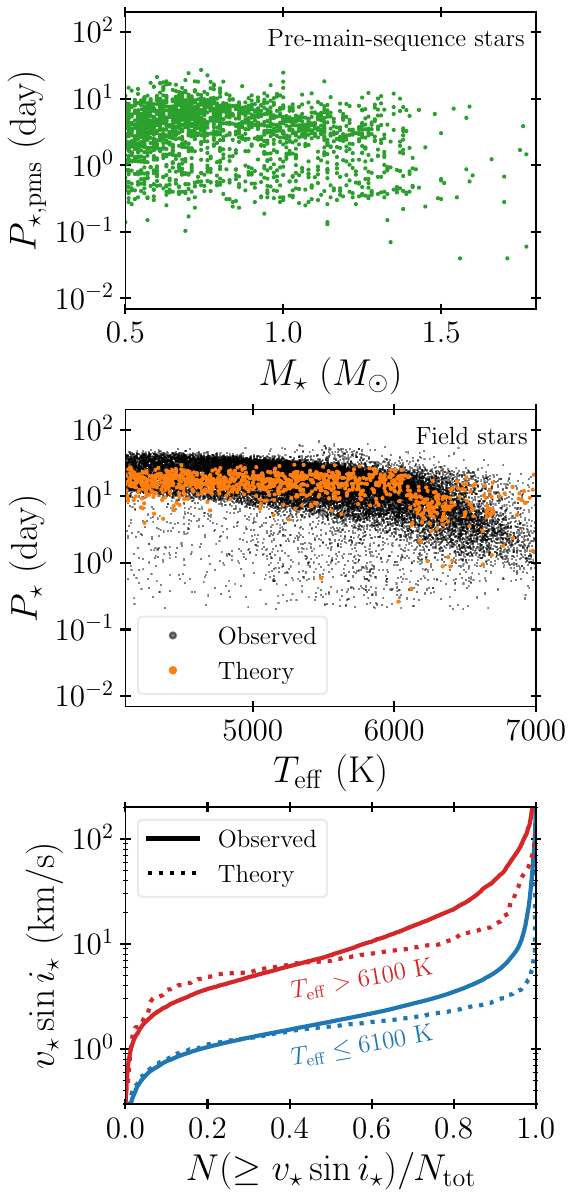}
\caption{
\textit{Top}: Rotation periods $P_{\star, \rm pms}$ and masses $M_\star$ of our sample of pre-main-sequence stars \citep{HendersonStassun(2012), Sinha+(2021), Getman+(2023)}.
\textit{Middle}: Observed \textit{Kepler} field star rotation periods $P_\star$ \citep{Lu+(2021)} compared to $10^3$ $P_\star$ draws from our magnetic braking model.
\textit{Bottom}: Cumulative distributions for the projected rotational velocities of cool ($T_{\rm eff} \le 6100 \ {\rm K}$) and hot ($T_{\rm eff} > 6100 \ {\rm K}$) \textit{Kepler} field stars, versus $10^3$ $v_\star \sin i_\star$ draws from our model.  Our treatment of stellar spin-down reproduces the lower $v_\star \sin i_\star$ values of cool host stars, but under-predicts rotation speeds for the most rapidly spinning hosts.
\label{fig:pms_rot_model}
}
\end{figure}

Pre-main sequence rotation periods $P_{\star, \rm pms}$ tend to lie between $\sim$1-10 days. 
We utilize $P_{\star, \rm pms}$ measurements of young stars in open clusters \citep{HendersonStassun(2012), Getman+(2023)} and star-forming regions \citep{Sinha+(2021)}, with mass $M_\star$ and age $t_{\rm pms}$ estimates from stellar isochrones (Fig.~\ref{fig:pms_rot_model}, top panel).  We associate each $M_\star$ measurement with a \texttt{MESA} model: measurements $\Ms \in [0.5, 0.7] \Msun$ are associated with a $0.6 \Msun$ \texttt{MESA} model, $\Ms \in [0.7, 0.9] \Msun$ with $0.8 \Msun$, $\Ms \in [0.9, 1.1]\Msun$ with $1.0 \Msun$, $\Ms \in [1.1, 1.3]\Msun$ with $1.2 \Msun$, $\Ms \in [1.3, 1.5]\Msun$ with $1.4 \Msun$, and $\Ms \in [1.5, 1.8] \Msun$ with $1.6 \Msun$.

These $P_{\star, \rm pms}$ and $t_{\rm pms}$ measurements are used as initial conditions for our Monte Carlo calculation of field main-sequence rotation periods (ignoring for now tidal torques from hot Jupiters).  We draw \{$M_\star, P_{\star, \rm pms}, t_{\rm pms}$\} values for a star in the observational sample, and start our integration of equation \eqref{eq:dJsdt_wind} at $t_{\rm start} = \max(10 \ {\rm Myr}, t_{\rm pms})$ (pre-main sequence stars are thought to have their spin periods fixed by circumstellar disk locking over the disk lifetime of $\sim$10 Myr).  Equation~\eqref{eq:dJsdt_wind}, together with the star's associated \texttt{MESA} model, is used to evolve the spin period, starting from $P_{\star, \rm pms}$, and integrating to a final $P_\star$ at $t_{\rm end}$. 
Values for $t_{\rm end}$ are drawn from \cite{Lu+(2021)}, who grouped \textit{Kepler} field stars based on their effective temperatures and rotation periods, and estimated ages using stellar velocity dispersions.  
We compare the results of our spin-down model with 26,921 \textit{Kepler} field star rotation periods from \cite{Lu+(2021)} in Figure~\ref{fig:pms_rot_model} (middle panel), including stars with temperatures $4100 \ {\rm K} \le T_{\rm eff} \le 7000 \ {\rm K}$, and removing likely sub-giants with ages $> 12$ Gyr or $\log g \le 3.9$ between $5000 \ {\rm K} \le T_{\rm eff} \le 6000 \ {\rm K}$.
Rotation periods are plotted against main-sequence effective temperatures $T_{\rm eff}$, which for our spin-down model are assigned by drawing $T_{\rm eff}$ from uniform intervals matched to a given \texttt{MESA} model's stellar mass ($[4100, 5000]{\rm K}$ for $0.6 \Msun$, $[5000, 5600] {\rm K}$ for $0.8\Msun$, $[5600, 6100] {\rm K}$ for $1.0 \Msun$, $[6100, 6400] {\rm K}$ for $1.2 \Msun$, $[6400, 6700] {\rm K}$ for $1.4 \Msun$, $[6700, 7000]{\rm K}$ for $1.6 \Msun$).  


We see from the middle panel of Fig.~\ref{fig:pms_rot_model} that our magnetic braking model gives rotation periods in rough agreement with field star observations.
The decline in $P_\star$ with increasing $T_{\rm eff}$ is somewhat smoother in the observations than in the model (recall the latter uses a discontinuous $\bar K$ across $1.2 M_\odot$). The model also has difficulty capturing the shortest period stars, both below and above the Kraft break. More sophisticated treatments of magnetic braking, which include how the metallicity affects the star's spin-down rate \citep[e.g.][]{Amard+(2019), Gossage+(2021)}, could alleviate these discrepancies.

We can also compute projected stellar rotation velocities $v_\star \sin i_\star$.  For each observed star in the \textit{Kepler} sample, we draw the star's $\cos i_\star$ from the uniform interval $[0,1]$, and use the measured radius $R_\star$ to calculate the equatorial velocity $v_\star = 2\pi R_\star/P_\star$.  Similarly, for our synthetic sample, we draw $\cos i_\star$ from $[0,1]$, and use the \texttt{MESA} model $R_\star$ to calculate $v_\star = 2\pi R_\star/P_\star$. Cumulative distributions for the observed and predicted $v_\star \sin i_\star$ values are displayed in the bottom panel of Figure~\ref{fig:pms_rot_model}.
The magnetic braking model qualitatively reproduces how cool stars spin slower than hot stars, 
but underpredicts the highest stellar rotation velocities by a factor of $\sim$2--3. This is the same problem noted above for $P_\star$.


\subsection{Obliquity and spin distributions of hot Jupiter systems} \label{subsec:oblique}

\begin{figure*}
\centering
\includegraphics[width=\linewidth]{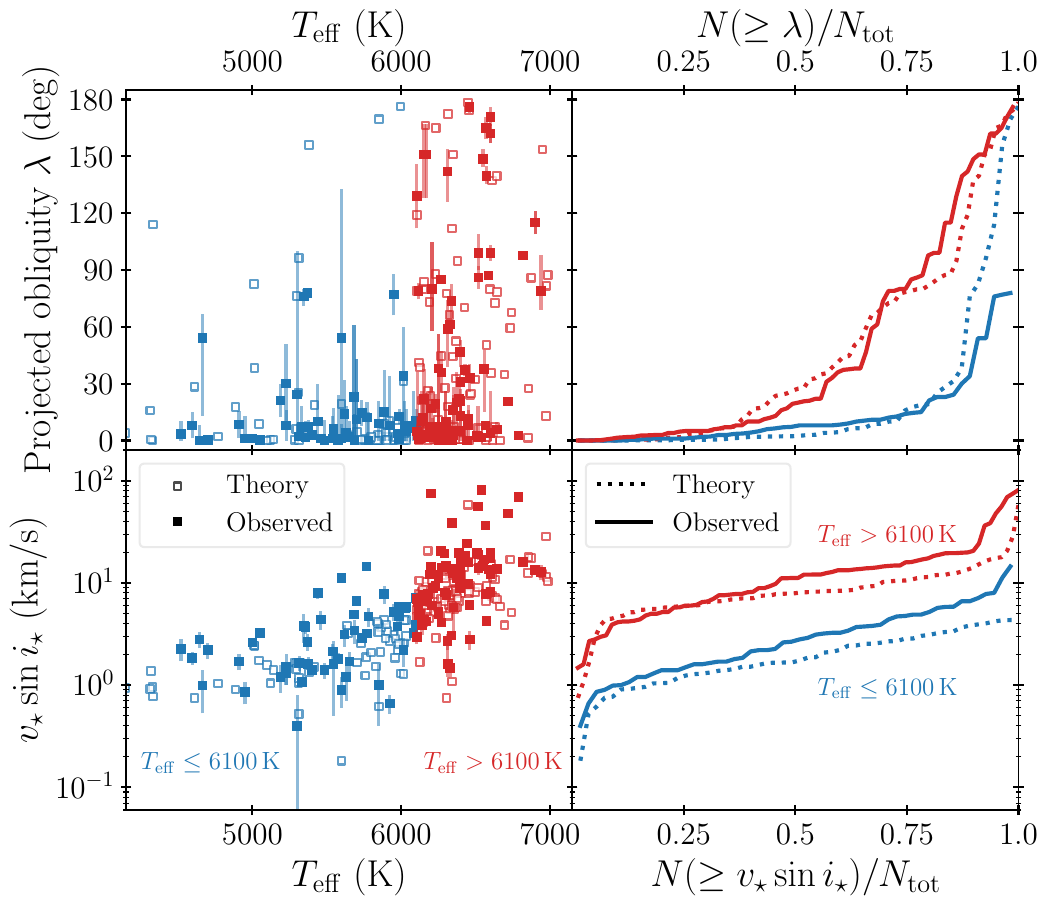}
\caption{
Projected obliquities $\lambda$ (top) and stellar rotation rates $v_\star \sin i_\star$ (bottom) vs.~host star effective temperature $T_{\rm eff}$, comparing predictions from our resonance locking + magnetic braking model to observations \citep{Albrecht+(2022), Rice+(2022), Siegel+(2023), Knudstrup+(2024), Wang+(2024)}, for hot Jupiter systems ($\Mp \ge 0.2 \ \Mjup$ and $a/R_\star \le 12$) around cool ($T_{\rm eff} \le 6100 \ {\rm K}$, blue) and hot ($T_{\rm eff} > 6100 \ {\rm K}$, red) hosts.  Left panels display the predicted $\lambda$ and $v_\star \sin i_\star$ values from our theory, as well as the observations.  Right panels plot cumulative distribution functions.  Resonance locking and magnetic braking appear capable of reproducing the low $\lambda$ and low $v_\star \sin i_\star$ values for cool stars hosting hot Jupiters.
\label{fig:spinoblpop_Teff}
}
\end{figure*}

\begin{figure*}
\centering
\includegraphics[width=\linewidth]{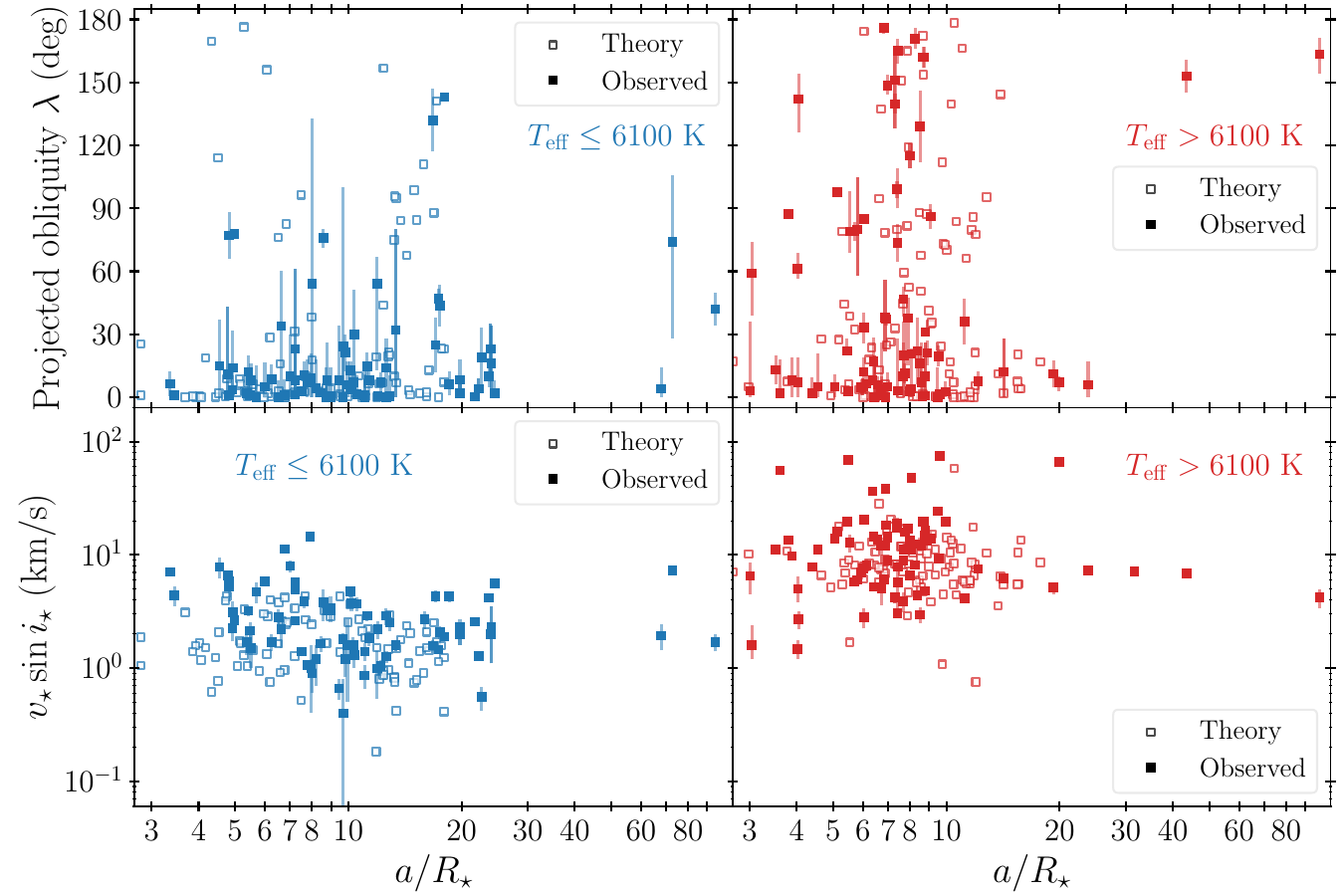}
\caption{
Projected stellar obliquities $\lambda$ (top) and projected stellar rotational velocities $v_\star \sin i_\star$ (bottom) vs.~hot Jupiter orbital distance $a/R_\star$, from both our resonance locking + magnetic braking population synthesis,  and observations \citep{Albrecht+(2022), Rice+(2022), Siegel+(2023), Knudstrup+(2024), Wang+(2024)}. Cool host stars are considered on the left, and hot host stars on the right. When $a/R_\star \lesssim 12$, hot stars are often misaligned, while cool stars are more aligned.  Observed rotational velocities of low-mass host stars appear to increase as $a/R_\star$ decreases, a trend that our model of tidal spin-up approximately reproduces.
\label{fig:spinoblpop_aRs}
}
\end{figure*}

Spin-orbit misalignments from high-eccentricity migration are not expected to depend on the host star's mass. We postulate that at the time of hot Jupiter formation, the obliquity distribution of low-mass hosts is identical to the observed distribution of high-mass hosts.  We draw projected obliquities $\lambda_0$ from observed hot host star systems ($6100 \ {\rm K} < T_{\rm eff} \le 7000 \ {\rm K}$, \citealt{Albrecht+(2022), Rice+(2022), Siegel+(2023), Knudstrup+(2024), Wang+(2024)}), and turn $\lambda_0$ into an initial 3D (``true'') obliquity $\psi_0$ using
\begin{equation}
    \tan \psi_0 \simeq \frac{\tan \lambda_0}{\sin \varphi_{\rm obs,0}},
\end{equation}
where we have approximated the orbit as edge-on (impact parameter $\simeq 0$),
with $\varphi_{\rm obs,0}$ drawn from the uniform interval $[0, \pi]$ \citep[e.g.][]{FabryckyWinn(2009)}.  We synthesize a population of  $10^4$ systems, drawing the planet's mass $\Mp$ and host's effective temperature $T_{\rm eff}$ in pairs $\{\Mp, T_{\rm eff}\}$ from observed systems with $\Mp \ge 0.2 \Mjup$ and $T_{\rm eff} \le 7000 \ {\rm K}$. 
We associate each system pair with a \texttt{MESA} model, whose mass depends on $T_{\rm eff}$ ($0.6 \Msun$ for $T_{\rm eff} \in [4100, 5000]{\rm K}$, $0.8\Msun$ for $[5000, 5600] {\rm K}$, $1.0 \Msun$ for $[5600, 6100] {\rm K}$, $1.2 \Msun$ for $[6100, 6400] {\rm K}$, $1.4 \Msun$ for $[6400, 6700] {\rm K}$, $1.6 \Msun$ for $[6700, 7000]{\rm K}$).  The star's $\{P_{\star, \rm pms}, t_{\rm pms}\}$ values are drawn from our sample of observed pre-main-sequence rotation periods (see section \ref{subsec:rot_vel} above), binned by the \texttt{MESA} model's mass (observed $\Ms \in [0.5, 0.7] \Msun$ for $0.6 \Msun$ \texttt{MESA} model, $\Ms \in [0.7, 0.9] \Msun$ for $0.8 \Msun$, $\Ms \in [0.9, 1.1]\Msun$ for $1.0 \Msun$, $\Ms \in [1.1, 1.3]\Msun$ for $1.2 \Msun$, $\Ms \in [1.3, 1.5]\Msun$ for $1.4 \Msun$, and $\Ms \in [1.5, 1.8] \Msun$ for $1.6 \Msun$).
The stellar spin is evolved from $t_{\rm start}$ to the time the hot Jupiter is assumed to form at $t_0 = 0.1 \ {\rm Gyr}$, using the magnetic braking model of section \ref{subsec:spin_down}.  The initial semi-major axis to stellar radius ratio $(a/\Rs)_0$ is drawn from a normal distribution with mean 13 and standard deviation 3.5.

After its assumed formation at $t_0$, the hot Jupiter may lock onto a resonance, causing the stellar obliquity and spin to evolve.  If the initial $(a/\Rs)_0 \le (a/\Rs)_{\rm crit} = 12$, we assume the dissipation rate is sufficiently strong to enforce a resonance lock (see eq.~\ref{eq:res_lock} of \citealt{Zanazzi+(2024)}).
In practice, the value of $(a/\Rs)_{\rm crit}$ likely depends on the mode amplitude, 
stratification profile, and how strongly the forced parent g-mode couples to child modes (via the three-mode coupling coefficient).  
The latter two quantities will vary with the mass and age of the star.
The viability of resonance locking could also be affected by so-called critical layers which form when the core's rotational frequency matches the mode pattern frequency \citep[e.g.][]{BookerBretherton1967, Hazel1967}.  See Section~\ref{sec:SummDisc} for further discussion.


We integrate equations~\eqref{eq:dota_lock}-\eqref{eq:dotpsi_lock}, picking mode azimuthal numbers $m$ according to the procedure laid out in section~\ref{sec:ResLock}.  We evolve each system until a time $t_{\rm end}$ drawn from the uniform interval $[1 \ {\rm Gyr}, t_{\rm max}]$,
with the lower bound of 1 Gyr motivated by the youngest age measurements of hot Jupiter hosts below $T_{\rm eff} \le 7000 \ {\rm K}$ \citep[e.g.][]{Albrecht+(2021), Albrecht+(2022)}.
At $t_{\rm end}$, our synthetic system is mock-observed to have a projected obliquity $\lambda$ and stellar inclination $i_\star$:
\begin{equation}
    \tan\lambda \simeq \tan \psi \sin\phi_{\rm obs},
    \hspace{4mm}
    \sin i_\star \simeq \frac{\cos \psi}{\cos \lambda},
\end{equation}
where the orbit is presumed edge-on (to yield a transit), 
with $\phi_{\rm obs}$ drawn from the uniform interval $[0, \pi]$ \citep[e.g.][]{FabryckyWinn(2009)}.  The final $P_\star$ and $R_\star$ values are converted into an observed projected velocity $v_\star \sin i_\star = (2\pi R_\star/P_\star) \sin i_\star$.  Any hot Jupiter with $a/\Rs < 2.7$ is assumed to tidally disrupt \citep[e.g.][]{Guillochon+(2011)} and removed from our synthetic population. About 11\% of the synthesized systems are so removed. 

Figures \ref{fig:spinoblpop_Teff} and \ref{fig:spinoblpop_aRs} display the results of our population synthesis. When plotting the synthetic data (``theory'') for these figures, we randomly pick 55 hot Jupiters ($a/R_\star \le 12$) around cool hosts ($T_{\rm eff} \le 6100 \ {\rm K}$), 79 hot Jupiters around hot hosts ($T_{\rm eff} > 6100 \ {\rm K}$), 8 warm Jupiters ($a/R_\star > 12$) around hot hosts, and 29 warm Jupiters around cool hosts --- this demographic breakdown matches the demographic breakdown of the 171 ``observed'' systems plotted in the figures. We find tidal alignment by resonance locking yields a $\lambda$-$T_{\rm eff}$ correlation similar to that observed. While $\lambda$ values remain primordially large for hot hosts, they are damped for cool hosts (Fig.~\ref{fig:spinoblpop_Teff}). 
This trend pertains to the closest hot Jupiter systems at $a/R_\star \lesssim 8$-12 (Fig.~\ref{fig:spinoblpop_aRs}). A discrepancy at these small separations is that our model predicts a somewhat higher proportion of high obliquities among low-mass stars ($\lesssim 10\%$ of the population). 
These higher obliquities tend to be seen in younger (ages $\lesssim 5$ Gyr),
lower-mass planets ($\Mp \lesssim 0.8 \Mjup$), which $m=-1$ and $m=-2$ locks have not had time to align (see Fig.~\ref{fig:SpinObl_retro}).


At larger separations ($a/R_\star \gtrsim 12$), resonances may be too weak to damp obliquities \citep{Zanazzi+(2024)}. Observed obliquities of such ``warm Jupiters'' in Fig.~\ref{fig:spinoblpop_aRs} indeed span a large range, around both hot and cool stars.
Some of these larger obliquities may be ascribed to stellar binary companions, 
e.g. HD 80606 b (the blue point at $a/R_\star = 95$ in Fig.~\ref{fig:spinoblpop_aRs}; \citealt{Herbard+(2010)}), K2-290 c (a red point at $a/R_\star = 44$; \citealt{Hjorth+(2021)}), and TIC 241249530 b (a red point at $a/R_\star = 98$; \citealt{Gupta+(2024)}); binary companions can tilt the orbits of warm Jupiters through Kozai oscillations after the disk dissipates \citep[e.g.][]{WuMurray(2003), FabryckyTremaine(2007), Vick+(2019), Vick+(2023)}, or through nodal precession of the protoplanetary disk before its dispersal \citep[e.g.][]{Batygin(2012), BatyginAdams(2013), ZanazziLai(2018), Gerbig+(2024)}.  
If we subtract off such binary systems in Fig.~\ref{fig:spinoblpop_aRs}, there is a suggestion that the remaining warm-Jupiter single hosts, both hot and cold,  
have mostly small obliquities (see also \citealt{Rice+(2022b), Wang+(2024)}).  Aligned warm Jupiters may have formed from disk migration rather than high-eccentricity migration.

Stellar rotational velocities $v_\star \sin i_\star$ are lower for cool stars than for hot stars from magnetic braking (Fig.~\ref{fig:spinoblpop_Teff}). Modeled velocities are systematically lower than observed velocities, a problem noted in our field-star tests that points to an inadequacy in our magnetic braking model (section \ref{subsec:rot_vel}). The effects of tidal spin-up are confined to low-mass stars hosting the closest hot Jupiters at $a/R_\star \lesssim 8-12$ (Fig.~\ref{fig:spinoblpop_aRs}).

\newpage
\section{Summary and Discussion}
\label{sec:SummDisc}

We have computed how a hot Jupiter locked in resonance with a gravity mode of its host star can affect stellar obliquity and spin rate. \cite{Zanazzi+(2024)} considered axisymmetric g-modes ($m=0$). Here we have considered non-axisymmetric ($m\ne 0$) modes. Non-axisymmetric modes affect not only stellar obliquity but also the magnitude of the stellar spin frequency. Our analysis accounts for spin changes from tidal torques, stellar evolution, and magnetic braking.

Our findings reinforce those of \cite{Zanazzi+(2024)}. Assuming the planet's potential with pattern frequency twice its mean motion ($k=2$) excites a stellar mode with angular degree $\ell = 2$, we found that resonant locks for all azimuthal wavenumbers $m = \{-2, -1, 0, 1, 2\}$ preferentially damp the obliquities of cool host stars as compared to hot host stars. Obliquities of cool stars damp more because their g-mode frequencies change more  on the main sequence; cool star g-modes are located in their radiative cores where hydrogen burning increases stratification over time, whereas hot star g-modes are confined to their radiative, non-burning envelopes. Cool host stars are also more amenable to being tidally torqued because their spin angular momenta are less than planetary orbital angular momenta; magnetic braking preferentially slows the convective envelopes of low-mass stars. While all $m < 2$ modes can damp low-mass stellar obliquities to zero within their main-sequence lifetimes, the $m=+2$ mode is relatively inefficient at damping, as can be seen in Fig.~\ref{fig:SpinObl_prog}, and by comparing $\tau_{+2}$ to the other dimensionless torques in equation \eqref{eq:tau_vals} as $\psi \rightarrow 0$. Planetary torques spin up host stars for $m > 0$, and spin down stars for $m < 0$; the effects can be comparable in magnitude to those of magnetic braking.

We performed a population synthesis of hot Jupiter systems to more directly compare with observed obliquities $\psi$ and stellar rotation velocities $v_\star \sin i_\star$. In the synthetic population, 
all starting obliquities $\psi_0$ are drawn from the same underlying random distribution, presumably the result of high-eccentricity migration (see discussion in Sec.~\ref{subsec:oblique}).
Each modeled hot Jupiter is assumed to start in resonant lock with a stellar g-mode having the azimuthal wavenumber $m$ that maximizes the mode energy for a given $\psi_0$. The higher the starting obliquity $\psi_0$, the lower the chosen $m$ (Fig.~\ref{fig:d2_amp}); e.g. $\psi_0 > 130^\circ$ is assumed to lock with an $m=-2$ mode, $\psi_0 \approx 90^\circ$ locks to $m=0$, and $\psi_0 < 50^\circ$ locks to $m=+2$. If and when $\psi$ drops to zero in the course of the evolution, the lock switches to $m=+2$, the only mode that can be excited for $\{k,\ell\}= \{2,2\}$. Initial spin periods are drawn from 
observations of young low-mass and high-mass stars; stellar properties are evolved with $\texttt{MESA}$; a simple magnetic braking model is adopted that is calibrated against the Sun; and mock observations of the projected obliquities $\lambda$ and projected rotation velocities $v_\star \sin i_\star$ are made for systems with ages drawn uniformly from 1 Gyr (similar to the youngest ages of our observed hot Jupiter sample) up to the main-sequence lifetime or 12 Gyr, whichever comes first.


Our population synthesis predicts correlations of obliquity with host star effective temperature (Fig.~\ref{fig:spinoblpop_Teff}) and planet separation (Fig.~\ref{fig:spinoblpop_aRs}) in broad agreement with observations. 
Perhaps the biggest discrepancy is that the theory still allows for a few retrograde, lower-mass hot Jupiters orbiting cool and relatively young stars; with planet masses $\lesssim 1 \, \Mjup$ and ages $\lesssim 5$ Gyr, these are not massive and old enough to have torqued their host stars into alignment. 
One possibility is that these retrograde hot Jupiters exist in nature, but orbit stars rotating too slowly to have their obliquities measured. Because $m=-2$ locks can slow stars down to periods $P_\star \gtrsim 30-50 \ {\rm d}$ (see Fig.~\ref{fig:SpinObl_retro}), a Rossiter-McLaughlin radial velocity signal may be difficult to detect \citep[e.g.][]{Albrecht+(2022)}.
Our retrograde hot Jupiters have projected rotational velocities $v_\star \sin i_\star$ in the range $\sim$0.3--1.5 km/s, lying at the lower edge of observed values (Figs.~\ref{fig:spinoblpop_Teff}-\ref{fig:spinoblpop_aRs}).  The lower masses of our apparently outlier retrograde hot Jupiters ($\sim$0.2--1.0 $\Mjup$) could further thwart a  $\lambda$ measurement.

Tidal spin-up causes $v_\star \sin i_\star$ to increase slightly with decreasing orbital distance, qualitatively reproducing the trend seen in observations (Fig.~\ref{fig:spinoblpop_aRs}). Our simplistic magnetic braking model de-spins cool stars 
a bit too strongly, under-predicting $v_\star \sin i_\star$ by factors of $\sim$2--3 (Figs.~\ref{fig:pms_rot_model}-\ref{fig:spinoblpop_aRs}). 

Recent works have claimed other obliquity correlations in hot Jupiter systems.  Around `moderately' hot host stars ($6100 \ {\rm K} \lesssim T_{\rm eff} \lesssim 7000 \ {\rm K}$), obliquities seem to drop for sufficiently massive hot Jupiters ($\Mp \gtrsim 13 M_{\rm Jup}$, e.g. \citealt{Herbrard+(2011), Albrecht+(2022), Giacalone+2024}).  The obliquities of subgiants with nascent radiative cores also tend to be low \citep[e.g.][]{Saunders+2024}.  It remains to be seen whether resonance locking can explain these emerging trends.

Stars become fully convective at masses $\Ms \lesssim 0.35 \Msun$ or effective temperatures $T_{\rm eff} \lesssim 3400 \ {\rm K}$; for these especially cool low-mass stars, core opacities become large and trigger convection (e.g.~\citealt{ChabrierBaraffe(1997)}; \citealt{Kippenhahn+(2013)}). 
Without a radiative zone, there are no g-modes for the planet to lock to, and stellar obliquities may remain at primordial values. So far, only one projected obliquity of $\lambda = 3.0^{+3.2}_{-3.7}$ deg has been measured for an M dwarf having $T_{\rm eff} = 3794 \ {\rm K}$  \citep{Gan+(2024)}.  If hot Jupiters around mid-to-late M-dwarf hosts form via high-eccentricity migration, resonance locking predicts obliquities should rise back up as $T_{\rm eff}$ drops below $\sim$3400 K.


A fundamental unresolved issue is whether resonance locking can even work in hot Jupiter systems. Resonance locking relies on the energy dissipation rate increasing as the tidal forcing frequency approaches the stellar mode oscillation frequency.  
Because of geometrical focusing in stellar cores, gravity wave amplitudes grow to large values, and non-linear effects alter the dissipation rate.
In particular, the spawning of `child' modes via a parametric instability has opened a host of issues  \citep[e.g.][]{KumarGoodman(1996), WuGoldreich(2001), BarkerOgilvie(2011), Weinberg+(2012), VanBeeck+(2024)}. 
\cite{EssickWeinberg(2016)} posit that child modes are standing waves that equilibrate with the tidal forcing to render the parent mode amplitude independent of the de-tuning frequency, thereby defeating resonance locking.  By contrast, \cite{Zanazzi+(2024)} argue that these child standing waves are of such large amplitude that they break, becoming traveling waves that restore the resonant response of the parent mode.  Neither  \cite{EssickWeinberg(2016)} nor \cite{Zanazzi+(2024)} accounted for `critical layers,' which absorb 
angular momentum from a gravity wave when the local rotation rate of the star matches the mode pattern frequency \citep[e.g.][]{BookerBretherton1967, Hazel1967}.  Critical layers develop when gravity modes preferentially spin up the star's core \citep[e.g.][]{BarkerOgilvie(2010), Barker(2011)}.  \cite{Guo+(2023)} found critical layers cause modes to become traveling waves, 
muting 
the frequency dependence of the tidal response close to resonance.  How 
the coupled modes from parametric instability are affected by a critical layer is unclear: does the layer absorb the child and restore the parent's resonant response, or does the layer absorb both parent and child and wipe out the resonance completely?  We hope to decide this debate with hydrodynamical simulations of gravity modes in stellar cores.


\vspace{0.2in}
\noindent 
We thank the referee, Adrian Barker, for their thoughtful comments which improved the quality of this manuscript. This work was supported by a 51 Pegasi b Heising-Simons Fellowship awarded to JJZ, and a Simons Investigator grant to EC.

\software{astropy \citep{Astropy_1,Astropy_2},  
          GYRE \citep{TownsendTeitler(2013), TownsendZweibel(2018), GoldsteinTownsend(2020)}
          MESA \citep{Paxton+(2011),Paxton+(2013),Paxton+(2015),Paxton+(2018),Paxton+(2019),Jermyn+(2023)}, 
          numpy \citep{numpy_cite},
          pandas \citep{pandas_cite},
          PyVista \citep{pyvista},
          scipy \citep{scipy_cite}
          }

\bibliography{main_v3}
\bibliographystyle{aasjournal}



\end{document}